\documentclass[12pt,draftcls, peerreview, onecolumn]{IEEEtran}
\usepackage[usenames]{color}
\usepackage{url}
\usepackage{cite}      % Written by Donald Arseneau
\usepackage{graphicx}  % Written by David Carlisle and Sebastian Rahtz
\usepackage{epsf}
\usepackage{epsfig}

\usepackage{subfigure} % Written by Steven Douglas Cochran
\usepackage{url}       % Written by Donald Arseneau
\usepackage{amsmath}   % From the American Mathematical Society
\makeatletter \newcommand{\vast}{\bBigg@{3}} \newcommand{\vvast}{\bBigg@{4}} \newcommand{\Vast}{\bBigg@{5}} \makeatother 
\DeclareMathAlphabet{\mathcalligra}{T1}{calligra}{m}{n}

\renewcommand{\baselinestretch}{2}

\begin{document}

% paper title
\title{
{Technical Report: Outage Performance of Full-Duplex MIMO DF Relaying using Zero-Forcing Beamforming}}
%\vspace{-1cm}

\author{Sung~Sik~Nam, Duckdong~Hwang, and Janghoon Yang
%Sung~Sik~Nam,~\IEEEmembership{Member,~IEEE,}
	
\thanks{
S.~S.~Nam is with Korea University, Korea (Email : ssnam@korea.ac.kr). Duckdong Hwang is with Konkuk University, Korea. (Email: duckdonh@yahoo.com). Janghoon Yang is with Seoul Media Institute of Technology, Korea.}
}

\markboth{S.S. Nam \MakeLowercase{\textit{et al.}}: FD-DF-ZFBF Arxiv}{Shell \MakeLowercase{\textit{et al.}}: FD-DF-ZFBF}

\maketitle
\vspace{-0.8cm}
\begin{abstract}
 In this paper, we deal with the performance analysis of full-duplex relaying in decode-\&-forward cooperative networks with multiple-antenna terminals. More specifically, by analyzing the end-to-end statistics, we derive the accurate closed-form expressions of the end-to-end outage probability for both transmit and receive ZFBF scheme over Rayleigh fading environments.
Some selected results show some interesting observations useful for system designers. Specifically, we observe that the outage performance can be improved by adopting the joint ZF-based precoding with different antenna configurations.
\end{abstract}

\vspace{-0.8cm}


\section{Introduction}  \label{intro}
Recently, research on cooperative communications with full-duplex (FD) relaying technique to achieve more efficient spectrum resource utilization, in addition to the network coverage extension and higher throughput, has underway as one of active research area \cite{Z.Zhang_2015,S.Goyal_2015}.
FD technique is considered as an essential component of the coming 5G wireless systems. In FD operation, the main drawback is the performance degradation imposed by the loopback self-interference due to signal leakage from the transmitter to the receiver at the relay side \cite{T.Riihonen_2011_2,B.P.Day_2012}.
Thus, the priority challenge issue is that of finding the suppression/cancellation techniques of the loopback self-interference. There were considered several approaches including combinations of analogue/digital self-interference (SI) cancellation  with RF domain approaches in addition to physical isolation between the transmit and receive antennas, for example, spatial-domain suppression, time-domain cancellation, zero-forcing (ZF) beamforming, power control and so on \cite{Z.Zhang_2015,Bharadia_2013,T.Riihonen_2011_2, T.Riihonen_2016,Suraweera_2014,G.Zheng_2015,Y.Li_2016}.

Especially, \cite{Suraweera_2014} has dealt with the deployment of FD relaying in amplify-\&-forward (AF) cooperative networks with multiple-antenna terminals. More specifically, the joint precoding/decoding design with the rank-1 ZF loopback self-interference suppression which maximizes the end-to-end performance has been proposed with the closed-form precoder/decoder solutions for transmit and receive ZF schemes and related closed-form performance results by solving appropriate optimization problems. In \cite{Suraweera_2014}, authors showed that with the proposed joint ZF-based precoding,  the end-to-end performance can be significantly improved.

There are two distinguished relaying protocols in the literature, such as AF and decode-\&-forward (DF). With the AF protocol, an amplified unwanted signal, such as not the interference and noise but also the loop interference in case of FD relaying, can be forwarded to the destination, leading to the noise amplification while the DF protocol provides more reliability for moderate to strong loop interferences by paying more processing compared to the AF protocol because with the DF protocol, the possible performance degradation due to the unwanted signal amplification can be mitigated by decoding the source message and then retransmitting the regenerated decoded message to the destination. 
Further, for a FD relay terminal, the optimal/suboptimal diversity-multiplexing tradeoff for this multi-hop setup is achievable by DF relaying \cite{D.Gunduz_2008}.

Based on above observations, in this paper, we consider the outage performance analysis of FD relaying cooperative networks with multiple-antenna terminals, especially in DF protocol, using ZF beamforming (ZFBF) by adopting the joint precoding/decoding design based on the AF cooperative relaying network in  \cite{Suraweera_2014}.
More specifically, by analyzing the end-to-end statistics, we derive the accurate closed-form expressions of the end-to-end outage probability for both transmit and receive ZFBF scheme over Rayleigh fading environments.

\section{System and Channel Models} \label{sec_1}
\begin{figure} 
\centering
\includegraphics[width=5in, trim=1.15cm 0cm 1.2cm 0cm, clip]{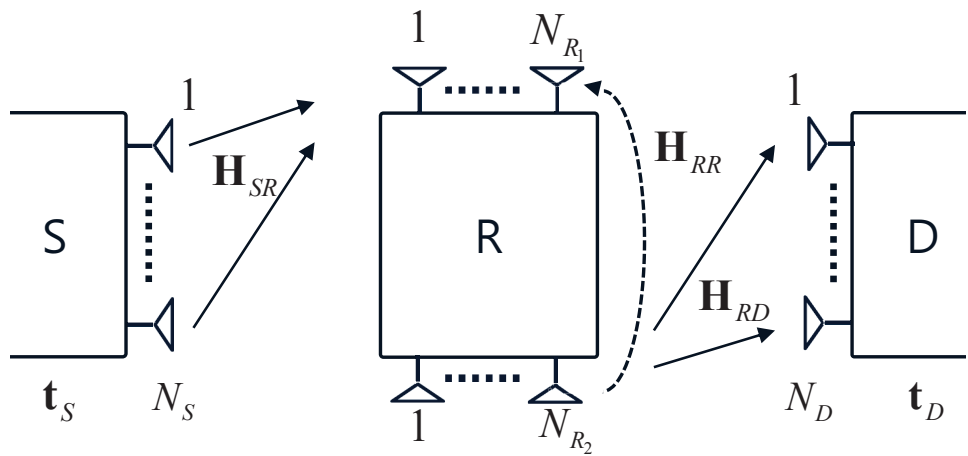}
\caption{System Model} \label{system_model}
\end{figure}

We employ a conventional three-node FD MIMO relay network with DF protocol consisting of one source (S), one relay (R), and
one destination (D), as shown in Fig.~\ref{system_model}. We assume that S has  $N_S$  antennas and D has  $N_D$  antennas while R is equipped with  $N_{R_2}$  transmit antennas and  $N_{R_1}$  receive antennas for FD operation. S has no direct link to D, which may result from heavy path loss and high shadowing between S and D.
We also assume that we adopt all known practical RF/analog domain interference cancellation approaches \cite{Bharadia_2013,S.Goyal_2015} to suppress the SI through the feedback channel.\footnote{This SI introduced by loop-back channel dominates over the intended reception and cannot be perfectly cancelled in practice. Note that according to recently published results \cite{Bharadia_2013,S.Goyal_2015}, the SI can be reduced up to a sufficient level (e.g., near 100 dB interference suppression).}
Further, we assume that a single data stream is transmitted. More specifically, S applies a precoding vector  ${\bf{t}}_S$  on the data stream, while D applies a linear receive vector  ${\mathbf{t}}_D$  with  $\left\| {{{\bf{t}}_S}} \right\|_F^2 = 1$  and  $\left\| {{{\bf{t}}_D}} \right\|_F^2 = 1$\normalsize, respectively, where  $\left\| \cdot \right\|_F$  denotes the Frobenius norm.

To suppress remained SI at a FD node, we apply conventional ZF self-interference suppression approaches at R node where ZFBF is an intuitive criterion.
Let, for the transmit beamforming scheme, the beamformer has the relay transmit vector, ${\bf{W}}_T$, at transmitter side of R with  $N_{R_2} \times 1$  and the relay receive vector,  ${\bf{W}}_R$, at receiver side of R with  $N_{R_1} \times 1$, then the SI can be written as  ${\bf{W}}_R^\dag {{\bf{H}}_{RR}}{{\bf{W}}_T}$. Here, ZF constraint is that this matrix product of
SI is forced into the all zero \cite{Suraweera_2014,D.Hwang_2016}, i.e.,  ${\bf{W}}_R^\dag {{\bf{H}}_{RR}}{{\bf{W}}_T} \!= \!0$, where all possible pairs  $\left({\bf{W}}_R, {\bf{W}}_T \right)$  satisfying the ZF constraint constitute the ZFBF solution set.

We denote that ${{\mathbf{{H}}}_{SR}}$ and ${{\mathbf{{H}}}_{RD}}$ are the S-R and R-D channels,
respectively, while ${{\mathbf{{H}}}_{RR}}$ denotes the loopback self-interference channel. 
We also assume that all the channels between nodes experience block fading. Thus, they remain constant over a long observation time (i.e., time slot), and varies independently from
one slot to another. This assumption applies to networks with a low mobility and corresponds to slow fading (block) channels where coding is performed over one block.
In addition, all links are subject to non-selective independent Rayleigh block fading
and additive white Gaussian noise (AWGN).

Then, let the equivalent S-R and R-D channels be ${{\bf{h}}_{SR}} = {\bf{{H}}}_{SR}{{\bf{t}}_S}$ and ${{\bf{h}}_{RD}} = {\bf{H}}_{RD}^\dag {{\bf{t}}_D}$, respectively, then the input signal at R and the received signal at R  can be written as respectively
\begin{equation} \label{sec_1:eq_1}
{{\mathbf{r}}_{IN}} = {{\bf{h}}_{SR}}{x_S} + {{\bf{H}}_{RR}}{{\bf{W}}_T}{x_R} + {{\bf{n}}_{RR}},
\end{equation}
and
\begin{equation} \label{sec_1:eq_2}
{r_R} = {\bf{W}}_R^\dag {\bf{r}} = {\bf{W}}_R^\dag {{\bf{h}}_{SR}}{x_S} + {\bf{W}}_R^\dag {{\bf{H}}_{RR}}{{\bf{W}}_T}{x_R} + {\bf{W}}_R^\dag {{\bf{n}}_{RR}},
\end{equation}
where ${x_S}$ is the transmitted symbol at S with zero-mean and average power $E\left\{ {{{\left| {{x_S}} \right|}^2}} \right\} = {P_S}$, , ${\bf{n}}_{RR}$ is ${N_{{R_1}}} \times 1$ AWGN vector with zero-mean and identity covariance matrix $E\left\{ {{{\bf{n}}_{RR}}{\bf{n}}_{RR}^\dag } \right\} = {{\bf{I}}_{{N_{{R_1}}}}}$, and $\left(\cdot \right)^\dag$ denotes the conjugate transpose.

Here, we add the ZF constraint that the design of ${\bf{W}}_T$ and ${\bf{W}}_R$ ensure no loopback self-interference for full-duplex operation at R. 
Then, once the ZF constraint is met, the received signal after ${\bf{W}}_R$ at R becomes as
\begin{equation} \label{sec_1:eq_4}
{r_R}^\prime  = {\bf{W}}_R^\dag {\bf{r}} = {\bf{W}}_R^\dag {{\bf{h}}_{SR}}{x_S} + {\bf{W}}_R^\dag {{\bf{n}}_{RR}},
\end{equation}
with the covariance and the received power at R as, respectively,
\begin{itemize}
\item {\textbf{Covariance: }} 
\begin{equation} \label{sec_1:eq_5}
\begin{aligned}
{\sum\nolimits_R} =& E\left\{ {{{\left| {{r_R}^\prime } \right|}^2}} \right\} = E\left\{ {\left( {{\bf{W}}_R^\dag {{\bf{h}}_{SR}}{x_S} + {\bf{W}}_R^\dag {{\bf{n}}_{RR}}} \right)\left( {x_S^*{\bf{h}}_{SR}^\dag {{\bf{W}}_R} + {\bf{n}}_{RR}^\dag {{\bf{W}}_R}} \right)} \right\}
\\
=& E\left\{ {{\bf{W}}_R^\dag {{\bf{h}}_{SR}}{x_S}x_S^*{\bf{h}}_{SR}^\dag {{\bf{W}}_R} + {\bf{W}}_R^\dag {{\bf{h}}_{SR}}{x_S}{\bf{n}}_{RR}^\dag {{\bf{W}}_R} + {\bf{W}}_R^\dag {{\bf{n}}_{RR}}x_S^*{\bf{h}}_{SR}^\dag {{\bf{W}}_R}} \right.
\\
&\quad\left.{+ {\bf{W}}_R^\dag {{\bf{n}}_{RR}}{\bf{n}}_{RR}^\dag {{\bf{W}}_R}} \right\}
\\
=& {\bf{W}}_R^\dag {{\bf{h}}_{SR}}E\left\{ {{{\left| {{x_S}} \right|}^2}} \right\}{\bf{h}}_{SR}^\dag {{\bf{W}}_R} + {\bf{W}}_R^\dag E\left\{ {{{\bf{n}}_{RR}}{\bf{n}}_{RR}^\dag } \right\}{{\bf{W}}_R}
\\
=& {P_S}{\bf{W}}_R^\dag {{\bf{h}}_{SR}}{\bf{h}}_{SR}^\dag {{\bf{W}}_R} + {\bf{W}}_R^\dag {{\bf{I}}_{{N_{{R_1}}}}}{{\bf{W}}_R}
\\
=& {P_S}{\bf{W}}_R^\dag {{\bf{h}}_{SR}}{\bf{h}}_{SR}^\dag {{\bf{W}}_R} + {\bf{W}}_R^\dag {{\bf{W}}_R},
\end{aligned}
\end{equation}
  \item {\textbf{Received Power: }}
\begin{equation} \label{sec_1:eq_6}
\begin{aligned}
{\rm Tr}\left( {{\sum\nolimits_R}} \right) =& {P_S}{\rm Tr}\left( {\left( {{\bf{W}}_R^\dag {{\bf{h}}_{SR}}} \right){{\left( {{\bf{W}}_R^\dag {{\bf{h}}_{SR}}} \right)}^\dag }} \right) + {\rm Tr}\left( {\left( {{\bf{W}}_R^\dag } \right){{\left( {{\bf{W}}_R^\dag } \right)}^\dag }} \right)\\
 =& {P_S}{\left| {{\bf{W}}_R^\dag {{\bf{h}}_{SR}}} \right|^2} + \left\| {{\bf{W}}_R^\dag } \right\|_F^2\\
 =& {P_S}{\left| {{\bf{W}}_R^\dag {{\bf{h}}_{SR}}} \right|^2} + 1 \mathop  = \limits^{{\rm{or}}} {P_S}{\left| {{\bf{h}}_{SR}^\dag {{\bf{W}}_R}} \right|^2} + 1,
\end{aligned}
\end{equation}
\end{itemize}
where ${\rm Tr}\left( \cdot \right)$ is the trace operation.

Then, for received signal at D, similarly, once the ZF constraint is met, the received signal at D can be given as
\begin{equation} \label{sec_1:eq_7}
{r_D} = {\bf{h}}_{RD}^\dag {{\bf{W}}_T}{x_R} + {n_{RD}},
\end{equation}
where ${x_R}$ is the relay transmit signal with zero-mean and average power $E\left\{ {{{\left| {{x_R}} \right|}^2}} \right\} = {P_R}$ and ${n_{RD}}$ is AWGN with zero mean and unit-variance.
Here, the covariance and the received power at D are as, respectively,
\begin{itemize}
  \item {\textbf{Covariance: }} 
\begin{equation} \label{sec_1:eq_8}
\begin{aligned}
{\sum\nolimits_D} =& E\left\{ {{{\left| {{r_D}} \right|}^2}} \right\} = E\left\{ {\left( {{\bf{h}}_{RD}^\dag {{\bf{W}}_T}{x_R} + {n_{RD}}} \right)\left( {x_R^*{\bf{W}}_T^\dag {{\bf{h}}_{RD}} + {n_{RD}}^*} \right)} \right\}\\
 =& E\left\{ {{\bf{h}}_{RD}^\dag {{\bf{W}}_T}{x_R}x_R^*{\bf{W}}_T^\dag {{\bf{h}}_{RD}} + {\bf{h}}_{RD}^\dag {{\bf{W}}_T}{x_R}{n_{RD}}^* + {n_{RD}}x_R^*{\bf{W}}_T^\dag {{\bf{h}}_{RD}} + {n_{RD}}{n_{RD}}^*} \right\}\\
 =& E\left\{ {{\bf{h}}_{RD}^\dag {{\bf{W}}_T}{x_R}x_R^*{\bf{W}}_T^\dag {{\bf{h}}_{RD}} + {n_{RD}}{n_{RD}}^*} \right\}\\
 =& {\bf{h}}_{RD}^\dag {{\bf{W}}_T}E\left\{ {{{\left| {{x_R}} \right|}^2}} \right\}{\bf{W}}_T^\dag {{\bf{h}}_{RD}} + E\left\{ {{n_{RD}}{n_{RD}}^*} \right\}\\
 =& {P_R}{\bf{h}}_{RD}^\dag {{\bf{W}}_T}{\bf{W}}_T^\dag {{\bf{h}}_{RD}} + 1,
\end{aligned}
\end{equation}
  \item {\textbf{Received Power: }}
\begin{equation}\label{sec_1:eq_9}
\begin{aligned}
{\rm Tr}\left( {{\sum\nolimits_D}} \right) =& {P_R} {\rm Tr}\left( {\left( {{\bf{h}}_{RD}^\dag {{\bf{W}}_T}} \right){{\left( {{\bf{h}}_{RD}^\dag {{\bf{W}}_T}} \right)}^\dag }} \right) + 1\\
 =& {P_R}{\left| {{\bf{h}}_{RD}^\dag {{\bf{W}}_T}} \right|^2} + 1 \mathop  = \limits^{{\rm{or}}} {P_R}{\left| {{\bf{W}}_T^\dag {{\bf{h}}_{RD}}} \right|^2} + 1.
\end{aligned}
\end{equation}
\end{itemize}
Note that as results with (\ref{sec_1:eq_4}) and (\ref{sec_1:eq_7}), we can adopt the design/analytical approaches used in \cite{Suraweera_2014}
 because our problem is eventually similar to what solve the problem of the joint precoding/decoding design in \cite{Suraweera_2014}.

\section{Statistical Analysis with Receive ZFBF}  \label{sec_2}
Based on \cite{K.Wong_2012,Suraweera_2014}, we first assume ${{\bf{W}}_T} = {{\bf{h}}_{RD}}$, then ${{\bf{W}}_R}$ should be aligned to the direction of ${{\bf{h}}_{SR}}$ projected to the orthogonal direction of ${{\bf{H}}_{RR}}{{\bf{h}}_{RD}}$. 
Therefore, by applying the similar approach used in \cite{Suraweera_2014}, ${{\bf{W}}_R}$ can be found from the projection onto the orthogonal space of ${{\bf{H}}_{RR}}{{\bf{h}}_{RD}}$, with the orthogonal projector onto the left null space of ${{\bf{H}}_{RR}}{{\bf{h}}_{RD}}$, ${\bf{\hat D}}$, such that
\begin{equation} \label{sec_2:eq_2}
{\bf{\hat D}} \buildrel \Delta \over = {{\bf{I}}_{{N_{{R_1}}}}} - \frac{{{{\bf{H}}_{RR}}{{\bf{h}}_{RD}}{{\left( {{{\bf{H}}_{RR}}{{\bf{h}}_{RD}}} \right)}^\dag }}}{{{{\left( {{{\bf{H}}_{RR}}{{\bf{h}}_{RD}}} \right)}^\dag }{{\bf{H}}_{RR}}{{\bf{h}}_{RD}}}} \mathop  = \limits^{{\rm{or}}} {{\bf{I}}_{{N_{{R_1}}}}} - \frac{{{{\bf{H}}_{RR}}{{\bf{h}}_{RD}}{\bf{h}}_{RD}^\dag {\bf{H}}_{RR}^\dag }}{{\left\| {{{\bf{H}}_{RR}}{{\bf{h}}_{RD}}} \right\|_F^2}},
\end{equation}
where
${\bf{\hat D}}$ is idempotent and $\frac{{{{\bf{H}}_{RR}}{{\bf{h}}_{RD}}{{\left( {{{\bf{H}}_{RR}}{{\bf{h}}_{RD}}} \right)}^\dag }}}{{{{\left( {{{\bf{H}}_{RR}}{{\bf{h}}_{RD}}} \right)}^\dag }{{\bf{H}}_{RR}}{{\bf{h}}_{RD}}}}$ is the orthogonal projector with rank one (isolate the signal in a single direction ${{\bf{H}}_{RR}}{{\bf{h}}_{RD}}$), and 
${{\bf{I}}_{{N_{{R_1}}}}} - \frac{{{{\bf{H}}_{RR}}{{\bf{h}}_{RD}}{{\left( {{{\bf{H}}_{RR}}{{\bf{h}}_{RD}}} \right)}^\dag }}}{{{{\left( {{{\bf{H}}_{RR}}{{\bf{h}}_{RD}}} \right)}^\dag }{{\bf{H}}_{RR}}{{\bf{h}}_{RD}}}}$ is the orthogonal projector with rank $\left( {{N_{{R_1}}} - 1} \right)$ (eliminate the signal in the direction ${{\bf{H}}_{RR}}{{\bf{h}}_{RD}}$) or it is also called the complementary projector to 
$\frac{{{{\bf{H}}_{RR}}{{\bf{h}}_{RD}}{{\left( {{{\bf{H}}_{RR}}{{\bf{h}}_{RD}}} \right)}^\dag }}}{{{{\left( {{{\bf{H}}_{RR}}{{\bf{h}}_{RD}}} \right)}^\dag }{{\bf{H}}_{RR}}{{\bf{h}}_{RD}}}}$. 

Once ${{\bf{W}}_R}$ is determined, the SNR monotonically depends on the quantity $\left\| {{\bf{\hat D}}{{\bf{h}}_{SR}}} \right\|_F^2$. 
Here, ${{\bf{t}}_S}$ is embedded in $\left\| {{\bf{\hat D}}{{\bf{h}}_{SR}}} \right\|_F^2$. 
Then, by adopting similar approach used in \cite{Suraweera_2014}, we can also consider that the SNR monotonically depends on the quantity $\left\| {{\bf{\hat D}}{{\bf{H}}_{SR}}} \right\|_F^2$ where $\left\| {{\bf{\hat D}}{{\bf{H}}_{SR}}} \right\|_F^2 = {\lambda _{\max }}\left( {{{\left( {{\bf{\hat D}}{{\bf{H}}_{SR}}} \right)}^\dag }\left( {{\bf{\hat D}}{{\bf{H}}_{SR}}} \right)} \right)={\lambda _{\max }}\left( {{\bf{H}}_{SR}^\dag {{{\bf{\hat D}}}^\dag }{\bf{\hat D}}{{\bf{H}}_{SR}}} \right)$.
Here, because ${\bf{\hat D}}$ is idempotent, $\left\| {{\bf{\hat D}}{{\bf{H}}_{SR}}} \right\|_F^2$ can be simplified as
\begin{equation} \label{sec_2:eq_3}
\left\| {{\bf{\hat D}}{{\bf{H}}_{SR}}} \right\|_F^2 = {\lambda _{\max }}\left( {{\bf{H}}_{SR}^\dag {\bf{\hat D}}{{\bf{H}}_{SR}}} \right).
\end{equation}
Therefore, we can re-write $\left\| {{\bf{\hat D}}{{\bf{H}}_{SR}}} \right\|_F^2$ as
\begin{equation} \label{sec_2:eq_4}
\left\| {{\bf{\hat D}}{{\bf{H}}_{SR}}} \right\|_F^2 = {\lambda _{\max }}\left( {{\bf{H}}_{SR}^\dag \left( {{{\bf{I}}_{{N_{{R_1}}}}} - \frac{{{{\bf{H}}_{RR}}{{\bf{h}}_{RD}}{{\left( {{{\bf{H}}_{RR}}{{\bf{h}}_{RD}}} \right)}^\dag }}}{{{{\left( {{{\bf{H}}_{RR}}{{\bf{h}}_{RD}}} \right)}^\dag }{{\bf{H}}_{RR}}{{\bf{h}}_{RD}}}}} \right){{\bf{H}}_{SR}}} \right).
\end{equation}
Here, $\frac{{{{\bf{H}}_{RR}}{{\bf{h}}_{RD}}{{\left( {{{\bf{H}}_{RR}}{{\bf{h}}_{RD}}} \right)}^\dag }}}{{{{\left( {{{\bf{H}}_{RR}}{{\bf{h}}_{RD}}} \right)}^\dag }{{\bf{H}}_{RR}}{{\bf{h}}_{RD}}}}$ has rank one. Thus, it can be re-written as
\begin{equation} \label{sec_2:eq_5}
\begin{aligned}
\left\| {{\bf{\hat D}}{{\bf{H}}_{SR}}} \right\|_F^2 =& {\lambda _{\max }}\left( {{\bf{H}}_{SR}^\dag {{\bf{U}}^\dag }\left( {{{\bf{I}}_{{N_{{R_1}}}}} - {\rm{diag}}\left( {1,0, \cdots ,0} \right)} \right){\bf{U}}{{\bf{H}}_{SR}}} \right)
\\
 =& {\lambda _{\max }}\left( {{\bf{\hat H}}_{SR}^\dag {\rm{diag}}\left( {0,1, \cdots ,1} \right){{{\bf{\hat H}}}_{SR}}} \right),
\end{aligned}
\end{equation}
where ${{{\bf{\hat H}}}_{SR}} = {\rm{ }}{\bf{U}}{{\bf{H}}_{SR}}$ and ${\bf{U}}$ is unitary matrix.
Thus, $\left\| {{\bf{\hat D}}{{\bf{H}}_{SR}}} \right\|_F^2$ can be finally re-written as
\begin{equation} \label{sec_2:eq_6}
\left\| {{\bf{\hat D}}{{\bf{H}}_{SR}}} \right\|_F^2 = {\lambda _{\max }}\left( {{\bf{\mathord{\buildrel{\lower3pt\hbox{$\scriptscriptstyle\smile$}} 
\over H} }}_{SR}^\dag {{{\bf{\mathord{\buildrel{\lower3pt\hbox{$\scriptscriptstyle\smile$}} 
\over H} }}}_{SR}}} \right),
\end{equation}
where ${{{\bf{\mathord{\buildrel{\lower3pt\hbox{$\scriptscriptstyle\smile$}} 
\over H} }}}_{SR}}$ is $\left( {{N_{{R_1}}} - 1} \right) \times {N_S}$ matrix.
Here, based on our channel model assumptions, ${\bf{\mathord{\buildrel{\lower3pt\hbox{$\scriptscriptstyle\smile$}} 
\over H} }}_{SR}^\dag {{{\bf{\mathord{\buildrel{\lower3pt\hbox{$\scriptscriptstyle\smile$}} 
\over H} }}}_{SR}}$ follows the Wishart distribution. As a result, $\left\| {{\bf{\hat D}}{{\bf{H}}_{SR}}} \right\|_F^2$ is the maximum eigenvalue of a Wishart matrix $\left( {{\bf{\mathord{\buildrel{\lower3pt\hbox{$\scriptscriptstyle\smile$}} 
\over H} }}_{SR}^\dag {{{\bf{\mathord{\buildrel{\lower3pt\hbox{$\scriptscriptstyle\smile$}} 
\over H} }}}_{SR}}} \right)$ with dimension $\left( {{N_{{R_1}}} - 1} \right) \times {N_S}$.

Then, the SNRs, ${\gamma _{SR,1}}$ and ${\gamma _{RD,1}}$, can be finally written as
\begin{equation} \label{sec_2:eq_7}
{\gamma _{SR,1}} = {\gamma _{SR\_\max ,1}} = {P_S}\left\| {{\bf{\hat D}}{{\bf{H}}_{SR}}} \right\|_F^2 = {P_S}{\lambda _{\max }}\left( {{\bf{\mathord{\buildrel{\lower3pt\hbox{$\scriptscriptstyle\smile$}} 
\over H} }}_{SR}^\dag {{{\bf{\mathord{\buildrel{\lower3pt\hbox{$\scriptscriptstyle\smile$}} 
\over H} }}}_{SR}}} \right) = {P_S}{\Lambda _{SR\_\max ,1}},
\end{equation}
and
\begin{equation} \label{sec_2:eq_8}
{\gamma _{RD,1}} = {\gamma _{RD\_\max ,1}} = {P_R}\left\| {{{\bf{H}}_{RD}}} \right\|_F^2 = {P_R}{\Lambda _{RD\_\max ,1}}.
\end{equation}
Here, based on channel assumptions in Sec.~\ref{sec_1}, the links are subject to i.i.d. Rayleigh block fading with the average SNRs, ${{\overline \gamma }_{SR}}$ and ${{\overline \gamma }_{RD}}$. It means that all the channels are stationary during a single transmission and identical.

Therefore, with the help of \cite{R.Mallik_2003}, the PDF expression of ${\Lambda _{SR\_\max ,1}}$ and ${\Lambda _{RD\_\max ,1}}$ can be written as
\begin{equation} \label{sec_2:eq_9}
{f_{{\Lambda _{SR\_\max ,1}}}}\left( \gamma  \right) = \sum\limits_{n = 1}^{\min \left( {{N_S},{N_{{R_1}}} - 1} \right)} {\sum\limits_{m = \left| {{N_S} - \left( {{N_{{R_1}}} - 1} \right)} \right|}^{\left( {{N_S} + {N_{{R_1}}} - 1} \right) \cdot n - 2{n^2}} {\frac{{D_{n,m}^1}}{{m!}}{{\left( {\frac{n}{{{{\overline \gamma }_{SR}}}}} \right)}^{\!\!\!m + 1}}{\gamma ^m}\exp \left( { - \frac{n}{{{{\overline \gamma }_{SR}}}} \cdot \gamma } \right)} },
\end{equation}
and
\begin{equation} \label{sec_2:eq_10}
{f_{{\Lambda _{RD\_\max ,1}}}}\left( \gamma  \right) = \sum\limits_{k = 1}^{\min \left( {{N_{{R_2}}},{N_D}} \right)} {\sum\limits_{l = \left| {{N_{{R_2}}} - {N_D}} \right|}^{\left( {{N_{{R_2}}} + {N_D}} \right) \cdot k - 2{k^2}} {\frac{{C_{k,l}^1}}{{l!}}{{\left( {\frac{k}{{{{\overline \gamma }_{RD}}}}} \right)}^{\!\!\!l + 1}}{\gamma ^l}\exp \left( { - \frac{k}{{{{\overline \gamma }_{RD}}}} \cdot \gamma } \right)} }.
\end{equation}
Then, by transforming density functions of (\ref{sec_2:eq_9}) and (\ref{sec_2:eq_10}) with (\ref{sec_2:eq_7}) and (\ref{sec_2:eq_8}), the PDF expressions of ${\gamma _{SR,1}}$ and ${\gamma _{RD,1}}$ can be finally obtained as
\begin{equation} \label{sec_2:eq_11}
\begin{aligned}
{f_{{\gamma _{SR,1}}}}\left( x \right) =& {f_{{\Lambda _{SR\_\max ,1}}}}\left( {\frac{x}{{{P_S}}}} \right) \cdot \frac{1}{{{P_S}}} 
\\
=& \sum\limits_{n = 1}^{\min \left( {{N_S},{N_{{R_1}}} - 1} \right)} {\sum\limits_{m = \left| {{N_S} - \left( {{N_{{R_1}}} - 1} \right)} \right|}^{\left( {{N_S} + {N_{{R_1}}} - 1} \right) \cdot n - 2{n^2}} {\frac{{D_{n,m}^1}}{{m!}}{{\left( {\frac{n}{{{P_S}{{\overline \gamma }_{SR}}}}} \right)}^{\!\!\!m + 1}}{x^m}\exp \left( { - \frac{n}{{{P_S}{{\overline \gamma }_{SR}}}} \cdot x} \right)} },
\end{aligned}
\end{equation}
and
\begin{equation} \label{sec_2:eq_12}
\begin{aligned}
{f_{{\gamma _{RD,1}}}}\left( x \right) =& {f_{{\Lambda _{RD\_\max ,1}}}}\left( {\frac{x}{{{P_R}}}} \right) \cdot \frac{1}{{{P_R}}} 
\\
=& \sum\limits_{k = 1}^{\min \left( {{N_{{R_2}}},{N_D}} \right)} {\sum\limits_{l = \left| {{N_{{R_2}}} - {N_D}} \right|}^{\left( {{N_{{R_2}}} + {N_D}} \right) \cdot k - 2{k^2}} {\frac{{C_{k,l}^1}}{{l!}}{{\left( {\frac{k}{{{P_R}{{\overline \gamma }_{RD}}}}} \right)}^{\!\!\!l + 1}}{x^l}\exp \left( { - \frac{k}{{{P_R}{{\overline \gamma }_{RD}}}} \cdot x} \right)} }.
\end{aligned}
\end{equation}
Here, $D_{n,m}^1 = {c_{n,m}}{K_{a,b}}\frac{{m!}}{{{n^{m + 1}}}}$ and $C_{k,l}^1 = {c_{k,l}}{K_{a,b}}\frac{{l!}}{{{k^{l + 1}}}}$ in \cite{R.Mallik_2003} where ${K_{a,b}} = \frac{1}{{\left[ {\prod\nolimits_{i = 1}^a {\left( {a - i} \right)!\left( {b - i} \right)!} } \right]}}$, $a = \min \left( {{N_S},{N_{{R_1}}} - 1} \right)$, and $b = \max \left( {{N_S},{N_{{R_1}}} - 1} \right)$ and the coefficients ${c_{n,m}}$ are determined by applying a curve fitting on the plot of $\frac{d}{{d\lambda }}a \times a$ Hankel matrix in \cite{R.Mallik_2003}.  Note that $D_{n,m}^1$ and $C_{k,l}^1$ can be computed with the help of \cite{R.Mallik_2003,S.Aissa_2005} by using most symbolic softwares such as Maple, Mathematica, or Matlab. In Appendix~\ref{AP:1}, for users' convenience, we provide the Matlab based code based on the proposed algorithm in \cite{S.Aissa_2005}. With this code, we can directly compute the exact values of the coefficients instead of any curve fitting or approximation.

\section{Statistical Analysis with Transmit ZFBF}  \label{sec_3}
In this case, similar to receive ZF case in Sec. \ref{sec_2}, by fixing ${{\bf{W}}_R} = {{\bf{h}}_{SR}}$,
${{\bf{W}}_T}$ should be aligned to the direction of ${{\bf{h}}_{RD}}$ projected to the orthogonal direction of ${\bf{H}}_{RR}^\dag {{\bf{h}}_{SR}}$. 
Therefore, ${{\bf{W}}_T}$ can be found from the projection onto the orthogonal space of ${\bf{H}}_{RR}^\dag {{\bf{h}}_{SR}}$ 
with the orthogonal projector onto the left null space of ${\bf{H}}_{RR}^\dag {{\bf{h}}_{SR}}$, ${\bf{\hat B}}$, such that 
\begin{equation} \label{sec_3:eq_3}
{\bf{\hat B}} \buildrel \Delta \over = {{\bf{I}}_{{N_{{R_2}}}}} - \frac{{{\bf{H}}_{RR}^\dag {{\bf{h}}_{SR}}{{\left( {{\bf{H}}_{RR}^\dag {{\bf{h}}_{SR}}} \right)}^\dag }}}{{{{\left( {{\bf{H}}_{RR}^\dag {{\bf{h}}_{SR}}} \right)}^\dag }{\bf{H}}_{RR}^\dag {{\bf{h}}_{SR}}}} \mathop  = \limits^{{\rm{or}}} {{\bf{I}}_{{N_{{R_2}}}}} - \frac{{{\bf{H}}_{RR}^\dag {{\bf{h}}_{SR}}{\bf{h}}_{SR}^\dag {{\bf{H}}_{RR}}}}{{\left\| {{\bf{H}}_{RR}^\dag {{\bf{h}}_{SR}}} \right\|_F^2}},
\end{equation}
where ${\bf{\hat B}}$ is idempotent, $\frac{{{\bf{H}}_{RR}^\dag {{\bf{h}}_{SR}}{{\left( {{\bf{H}}_{RR}^\dag {{\bf{h}}_{SR}}} \right)}^\dag }}}{{{{\left( {{\bf{H}}_{RR}^\dag {{\bf{h}}_{SR}}} \right)}^\dag }{\bf{H}}_{RR}^\dag {{\bf{h}}_{SR}}}}$ is the orthogonal projector with rank one, and ${{\bf{I}}_{{N_{{R_2}}}}} - \frac{{{\bf{H}}_{RR}^\dag {{\bf{h}}_{SR}}{{\left( {{\bf{H}}_{RR}^\dag {{\bf{h}}_{SR}}} \right)}^\dag }}}{{{{\left( {{\bf{H}}_{RR}^\dag {{\bf{h}}_{SR}}} \right)}^\dag }{\bf{H}}_{RR}^\dag {{\bf{h}}_{SR}}}}$ is the orthogonal projector with rank $\left( {{N_{{R_2}}} - 1} \right)$.
Then, similarly, ${{\bf{W}}_T}$ is determined, we can also consider that the SNR monotonically depends on the quantity $\left\| {{\bf{\hat B}}{{\bf{H}}_{RD}}} \right\|_F^2$, where $\left\| {{\bf{\hat B}}{{\bf{H}}_{RD}}} \right\|_F^2 = {\lambda _{\max }}\left( {{\bf{H}}_{RD}^\dag {\bf{\hat B}}{{\bf{H}}_{RD}}} \right)$. Then, $\left\| {{\bf{\hat B}}{{\bf{H}}_{RD}}} \right\|_F^2$ becomes
\begin{equation} \label{sec_3:eq_4}
\begin{aligned}
\left\| {{\bf{\hat B}}{{\bf{H}}_{RD}}} \right\|_F^2 =& {\lambda _{\max }}\left( {{\bf{H}}_{RD}^\dag {{\bf{U}}^\dag }\left( {{{\bf{I}}_{{N_{{R_2}}}}} - {\rm{diag}}\left( {1,0, \cdots ,0} \right)} \right){\bf{U}}{{\bf{H}}_{RD}}} \right)
\\
=& {\lambda _{\max }}\left( {{\bf{\hat H}}_{RD}^\dag {\rm{diag}}\left( {0,1, \cdots ,1} \right){{{\bf{\hat H}}}_{RD}}} \right),
\end{aligned}
\end{equation}
where ${{{\bf{\hat H}}}_{RD}} = {\bf{U}}{{\bf{H}}_{RD}}$ and ${\bf{U}}$ is unitary matrix. 
Then, $\left\| {{\bf{\hat B}}{{\bf{H}}_{RD}}} \right\|_F^2$ can be finally re-written as
\begin{equation} \label{sec_3:eq_5}
\left\| {{\bf{\hat B}}{{\bf{H}}_{RD}}} \right\|_F^2 = {\lambda _{\max }}\left( {{\bf{\mathord{\buildrel{\lower3pt\hbox{$\scriptscriptstyle\smile$}} 
\over H} }}_{RD}^\dag {{{\bf{\mathord{\buildrel{\lower3pt\hbox{$\scriptscriptstyle\smile$}} 
\over H} }}}_{RD}}} \right),
\end{equation}
where ${{{\bf{\mathord{\buildrel{\lower3pt\hbox{$\scriptscriptstyle\smile$}}\over H} }}}_{RD}}$ is $\left( {{N_{{R_2}}} - 1} \right) \times {N_D}$ matrix.
As a result, $\left\| {{\bf{\hat B}}{{\bf{H}}_{RD}}} \right\|_F^2$ is the maximum eigenvalue of a Wishart matrix $\left( {{\bf{\mathord{\buildrel{\lower3pt\hbox{$\scriptscriptstyle\smile$}} 
\over H} }}_{RD}^\dag {{{\bf{\mathord{\buildrel{\lower3pt\hbox{$\scriptscriptstyle\smile$}} 
\over H} }}}_{RD}}} \right)$ with dimension $\left( {{N_{{R_2}}} - 1} \right) \times {N_D}$. 
Then, similar to receive ZF case, the SNRs, ${\gamma _{SR,1}}$ and ${\gamma _{RD,1}}$, can be finally written as
\begin{equation} \label{sec_3:eq_6}
{\gamma _{SR,2}} = {\gamma _{SR\_\max ,2}} = {P_S}\left\| {{{\bf{H}}_{SR}}} \right\|_F^2 = {P_S}{\Lambda _{SR\_\max ,2}}\quad{\text{ with dimension }}\left( {{N_{{R_1}}} \times {N_S}} \right),
\end{equation}
\begin{equation} \label{sec_3:eq_7}
{\gamma _{RD,2}} = {\gamma _{RD\_\max ,2}} = {P_R}\left\| {{\bf{\hat B}}{{\bf{H}}_{RD}}} \right\|_F^2 = {P_R}{\lambda _{\max }}\left( {{\bf{\mathord{\buildrel{\lower3pt\hbox{$\scriptscriptstyle\smile$}} 
\over H} }}_{RD}^\dag {{{\bf{\mathord{\buildrel{\lower3pt\hbox{$\scriptscriptstyle\smile$}} 
\over H} }}}_{RD}}} \right) = {P_R}{\Lambda _{RD\_\max ,2}}.
\end{equation}

Therefore, similar to receive ZF case, the PDF expressions of ${\Lambda _{SR\_\max ,2}}$ and ${\Lambda _{RD\_\max ,2}}$ can be also written as
\begin{equation} \label{sec_3:eq_8}
{f_{{\Lambda _{SR\_\max ,2}}}}\left( \gamma  \right) = \sum\limits_{n = 1}^{\min \left( {{N_S},{N_{{R_1}}}} \right)} {\sum\limits_{m = \left| {{N_S} - {N_{{R_1}}}} \right|}^{\left( {{N_S} + {N_{{R_1}}}} \right) \cdot n - 2{n^2}} {\frac{{D_{n,m}^2}}{{m!}}{{\left( {\frac{n}{{{{\overline \gamma }_{SR}}}}} \right)}^{\!\!\!m + 1}}{\gamma ^m}\exp \left( { - \frac{n}{{{{\overline \gamma }_{SR}}}} \cdot \gamma } \right)} },
\end{equation}
and
\begin{equation} \label{sec_3:eq_9}
{f_{{\Lambda _{RD\_\max ,2}}}}\left( \gamma  \right) = \sum\limits_{k = 1}^{\min \left( {{N_D},{N_{{R_2}}} - 1} \right)} {\sum\limits_{l = \left| {{N_D} - \left( {{N_{{R_2}}} - 1} \right)} \right|}^{\left( {{N_D} + {N_{{R_2}}} - 1} \right) \cdot k - 2{k^2}} {\frac{{C_{k,l}^2}}{{l!}}{{\left( {\frac{k}{{{{\overline \gamma }_{RD}}}}} \right)}^{\!\!\!l + 1}}{\gamma ^l}\exp \left( { - \frac{k}{{{{\overline \gamma }_{RD}}}} \cdot \gamma } \right)} }.
\end{equation}
Then, similarly, with ${\gamma _{SR,2}} = {P_S}{\Lambda _{SR\_\max ,2}}$ and ${\gamma _{RD,2}} = {P_R}{\Lambda _{RD\_\max ,2}}$, the closed-form expressions of PDF of ${\gamma _{SR,2}}$ and ${\gamma _{RD,2}}$ can be finally obtained as
\begin{equation} \label{sec_3:eq_10}
\begin{aligned}
{f_{{\gamma _{SR,2}}}}\left( x \right) =& {f_{{\Lambda _{SR\_\max ,1}}}}\left( {\frac{x}{{{P_S}}}} \right) \cdot \frac{1}{{{P_S}}} 
\\
=& \sum\limits_{n = 1}^{\min \left( {{N_S},{N_{{R_1}}}} \right)} {\sum\limits_{m = \left| {{N_S} - {N_{{R_1}}}} \right|}^{\left( {{N_S} + {N_{{R_1}}}} \right) \cdot n - 2{n^2}} {\frac{{D_{n,m}^2}}{{m!}}{{\left( {\frac{n}{{{P_S}{{\overline \gamma }_{SR}}}}} \right)}^{\!\!\!m + 1}}{x^m}\exp \left( { - \frac{n}{{{P_S}{{\overline \gamma }_{SR}}}} \cdot x} \right)} },
\end{aligned}
\end{equation}
and
\begin{equation} \label{sec_3:eq_11}
\begin{aligned}
{f_{{\gamma _{RD,2}}}}\left( x \right) =& {f_{{\Lambda _{RD\_\max ,1}}}}\left( {\frac{x}{{{P_R}}}} \right) \cdot \frac{1}{{{P_R}}} 
\\
=& \sum\limits_{k = 1}^{\min \left( {{N_D},{N_{{R_2}}} - 1} \right)} {\sum\limits_{l = \left| {{N_D} - \left( {{N_{{R_2}}} - 1} \right)} \right|}^{\left( {{N_D} + {N_{{R_2}}} - 1} \right) \cdot k - 2{k^2}} {\frac{{C_{k,l}^2}}{{l!}}{{\left( {\frac{k}{{{P_R}{{\overline \gamma }_{RD}}}}} \right)}^{\!\!\!l + 1}}{x^l}\exp \left( { - \frac{k}{{{P_R}{{\overline \gamma }_{RD}}}} \cdot x} \right)} }.
\end{aligned}
\end{equation}

\section{Outage Performance} \label{sec_4}
The outage probability, ${{\mathop{\rm P}\nolimits} _{OUT}}$, is defined as the probability that the instantaneous end-to-end SNR 
falls below a target SNR. Here, based on the mode of operation in Sec.~\ref{sec_1}, the overall system outage occurs a communication  failure in one of two links 
(i.e., from S to R or from R to D). Therefore, the overall channel outage probability can be expressed in terms of individual link outage probabilities as
\begin{equation} \label{sec_4:eq_1}
{{\mathop{\rm P}\nolimits} _{OUT}} = {{\mathop{\rm P}\nolimits} _{OUT,SR}} + \left( {1 - {{\mathop{\rm P}\nolimits} _{OUT,SR}}} \right){{\mathop{\rm P}\nolimits} _{OUT,RD}},
\end{equation}
where
\begin{equation} \label{sec_4:eq_2}
{{\mathop{\rm P}\nolimits} _{OUT,SR}} = \Pr \left[ {{\gamma _{SR,i}} < {\gamma _T}} \right] = \int_0^{{\gamma _T}} {{f_{{\gamma _{SR,i}}}}\left( x \right)dx} \quad {\text{for  }}i = 1,2
\end{equation}
and
\begin{equation} \label{sec_4:eq_3}
{{\mathop{\rm P}\nolimits} _{OUT,RD}} = \Pr \left[ {{\gamma _{RD,i}} < {\gamma _T}} \right] = \int_0^{{\gamma _T}} {{f_{{\gamma _{RD,i}}}}\left( x \right)dx} \quad {\text{for  }}i = 1,2.
\end{equation}
Similarly, in terms of the mutual-information rate, the event of an information-outage occurs when the received date rate falls below some fixed threshold, ${R_0}$, and each outage probability terms can be expressed for $i=1,2$ as
\begin{equation} \label{sec_4:eq_4}
\begin{aligned}
{{\mathop{\rm P}\nolimits} _{OUT,SR}} =& \Pr \left[ {{{\log }_2}\left( {1 + {\gamma _{SR,i}}} \right) < {R_0}} \right] = \Pr \left[ {{\gamma _{SR,i}} < {2^{{R_0} } - 1}} \right]
\\
 =& \int_0^{{2^{{R_0} } - 1}} {{f_{{\gamma _{SR,i}}}}\left( x \right)dx},
\end{aligned}
\end{equation}
\begin{equation} \label{sec_4:eq_5}
\begin{aligned}
{{\mathop{\rm P}\nolimits} _{OUT,RD}} =& \Pr \left[ {{{\log }_2}\left( {1 + {\gamma _{RD,i}}} \right) < {R_0}} \right] = \Pr \left[ {{\gamma _{RD,i}} < {2^{{R_0}} - 1}} \right]
\\
=& \int_0^{{2^{{R_0}} - 1}} {{f_{{\gamma _{RD,i}}}}\left( x \right)dx}.
\end{aligned}
\end{equation}
Note that in this case, the closed-form expression of the outage probability can be directly obtained by replacing ${\gamma _T}$ in the closed-form result of (\ref{sec_4:eq_1}) with ${2^{{R_0}}-1}$.

\subsection{Closed-form results for Receive ZFBF case}  \label{sec_4_1}
For S-R link, by substituting the PDF in (\ref{sec_2:eq_11}) into the outage probability in (\ref{sec_4:eq_2}), the outage probability of S-R link can be written as
\begin{equation} \label{sec_4_1:eq_1}
\begin{aligned}
{{\mathop{\rm P}\nolimits} _{OUT,SR}} =& \int_0^{{\gamma _T}} {{f_{{\gamma _{SR,1}}}}\left( x \right)dx}
\\
=& \sum\limits_{n = 1}^{\min \left( {{N_S},{N_{{R_1}}} - 1} \right)} {\sum\limits_{m = \left| {{N_S} - \left( {{N_{{R_1}}} - 1} \right)} \right|}^{\left( {{N_S} + {N_{{R_1}}} - 1} \right) \cdot n - 2{n^2}} {\frac{{D_{n,m}^1}}{{m!}}{{\left( {\frac{n}{{{P_S}{{\overline \gamma }_{SR}}}}} \right)}^{m + 1}}\int_0^{{\gamma _T}} {{x^m}\exp \left( { - \frac{n}{{{P_S}{{\overline \gamma }_{SR}}}} \cdot x} \right)dx} } }.
\end{aligned}
\end{equation}
Then, by applying \cite[Eq. (3.381-1)]{kn:gradshteyn_6} and then mathematical simplification, the closed-form result can be obtained as the incomplete Gamma function
\begin{equation} \label{sec_4_1:eq_2}
{{\mathop{\rm P}\nolimits} _{OUT,SR}} = \sum\limits_{n = 1}^{\min \left( {{N_S},{N_{{R_1}}} - 1} \right)} {\sum\limits_{m = \left| {{N_S} - \left( {{N_{{R_1}}} - 1} \right)} \right|}^{\left( {{N_S} + {N_{{R_1}}} - 1} \right) \cdot n - 2{n^2}} {\frac{{D_{n,m}^1}}{{m!}}\cdot\gamma \!\left( {m + 1,\frac{{n{\gamma _T}}}{{{P_S}{{\overline \gamma }_{SR}}}}} \right)} },
\end{equation}
where $\gamma \left( { \cdot , \cdot } \right)$ is the lower incomplete Gamma function.

For R-D link, similarly, the outage probability of R-D link  (\ref{sec_4:eq_3}) can be written with the PDF (\ref{sec_2:eq_12}) as
\begin{equation} \label{sec_4_1:eq_3}
\begin{aligned}
{{\mathop{\rm P}\nolimits} _{OUT,RD}} =& \int_0^{{\gamma _T}} {{f_{{\gamma _{RD,1}}}}\left( x \right)dx} 
\\
 =& \sum\limits_{k = 1}^{\min \left( {{N_{{R_2}}},{N_D}} \right)} {\sum\limits_{l = \left| {{N_{{R_2}}} - {N_D}} \right|}^{\left( {{N_{{R_2}}} + {N_D}} \right) \cdot k - 2{k^2}} {\frac{{C_{k,l}^1}}{{l!}}{{\left( {\frac{k}{{{P_R}{{\overline \gamma }_{RD}}}}} \right)}^{\!\!\!l + 1}}\int_0^{{\gamma _T}} {{x^l}\exp \left( { - \frac{k}{{{P_R}{{\overline \gamma }_{RD}}}} \cdot x} \right)dx} } }.
\end{aligned}
\end{equation}
Then, with (3.381.1), the closed-form result can be also obtained as the incomplete Gamma function
\begin{equation} \label{sec_4_1:eq_4}
{{\mathop{\rm P}\nolimits} _{OUT,RD}} = \sum\limits_{k = 1}^{\min \left( {{N_{{R_2}}},{N_D}} \right)} {\sum\limits_{l = \left| {{N_{{R_2}}} - {N_D}} \right|}^{\left( {{N_{{R_2}}} + {N_D}} \right) \cdot k - 2{k^2}} {\frac{{C_{k,l}^1}}{{l!}}\cdot\gamma \!\left( {l + 1,\frac{{k{\gamma _T}}}{{{P_R}{{\overline \gamma }_{RD}}}}} \right)} }.
\end{equation}
Note that based on \cite{Q.Zhou_2006}, we can see that a diversity order of $N_S\left(N_{R_1}-1\right)$ and $N_{R_2}N_D$ at the receiver side and the transmitter side, respectively. Thus, the full-duplex receive ZF design can achieve a diversity order of $\min\left(N_S\left(N_{R_1}-1\right),N_{R_2}N_D\right)$.

\subsection{Closed-form results for Transmit ZFBF case}  \label{sec_4_2}
In this case, the outage probability formulas have the similar integral form. Therefore, with the closed-form results of PDFs in (\ref{sec_3:eq_10}) and (\ref{sec_3:eq_11}), by adopting the same definite integral table used in (\ref{sec_4_1:eq_2}) and (\ref{sec_4_1:eq_4}),
both closed-form results can be obtained, respectively as
\begin{eqnarray}
\!\!\!\!\!\!\!\!\!\!\!\!&&{{\mathop{\rm P}\nolimits} _{OUT,SR}}\! =\! \int_0^{{\gamma _T}} {{f_{{\gamma _{SR,2}}}}\left( x \right)dx}  \!= \!\sum\limits_{n = 1}^{\min \left( {{N_S},{N_{{R_1}}}} \right)} {\sum\limits_{m = \left| {{N_S} - {N_{{R_1}}}} \right|}^{\left( {{N_S} + {N_{{R_1}}}} \right) \cdot n - 2{n^2}} {\frac{{D_{n,m}^2}}{{m!}}\cdot\gamma \!\left( {m + 1,\frac{{n{\gamma _T}}}{{{P_S}{{\overline \gamma }_{SR}}}}} \right)} },
\\
\!\!\!\!\!\!\!\!\!\!\!\!&&{{\mathop{\rm P}\nolimits} _{OUT,RD}}\! =\! \int_0^{{\gamma _T}} {{f_{{\gamma _{RD,2}}}}\left( x \right)dx}  \!=\! \sum\limits_{k = 1}^{\min \left( {N_D},{{N_{{R_2}}} - 1} \right)} {\sum\limits_{l = \left| {{N_D}-\left( {{N_{{R_2}}} - 1} \right)} \right|}^{\left( {\left({N_D}+{N_{{R_2}}} - 1\right)} \right) \cdot k - 2{k^2}} {\frac{{C_{k,l}^2}}{{l!}}\cdot\gamma \!\left( {l + 1,\frac{{k{\gamma _T}}}{{{P_R}{{\overline \gamma }_{RD}}}}} \right)} }.
\end{eqnarray}
Note also that, similarly, we can see that a diversity order of $N_D\left(N_{R_2}-1\right)$ and $N_S N_{R_1}$ at the transmitter side and the receiver side, respectively. Thus, the full-duplex transmit ZF design can achieve a diversity order of $\min\left(N_D\left(N_{R_2}-1\right),N_S N_{R_1}\right)$.

\section{Results and Conclusions} \label{conc}
In this section, some selected results for the outage probability are given. The simulation set-up follows the system model provided in Section II, especially  $P_S=P_R$  and  $\overline\gamma_{SR}=\overline\gamma_{RD}$ . Although we have mainly considered a symmetric setup, we additionally consider the effect of asymmetric setups. 
More specifically, we consider both cases; i) when the first
hop dominates over the second hop, e.g.,  ${P'_S}(={\alpha_{SR}}^{2} P_S)> {P'_R}(={\alpha_{RD}}^{2} P_R)$ , and ii) when the second hop dominates over the first hop, e.g.,  ${P'_R}(={\alpha_{RD}}^{2} P_R)> {P'_S}(={\alpha_{SR}}^{2} P_S)$ , where  $\alpha_{SR}$  and  $\alpha_{RD}$  are the path-loss factor for the S-R link and the R-D link, respectively.
In addition,  $(N_S,N_{R_1},N_{R_2},N_D)$  represents the each antenna configuration, where  $N_S$ ,  $N_D$ ,  $N_{R_1}$ , and  $N_{R_2}$  are the number of antennas at S, D, receiver side of R, and transmitter side of R, respectively. Further, in the following figures, the lines and the markers represent the simulation and the analytical results, respectively. Note that the simulation results match the derived analytical results well.

\begin{figure} 
\centering
\includegraphics[width=7in]{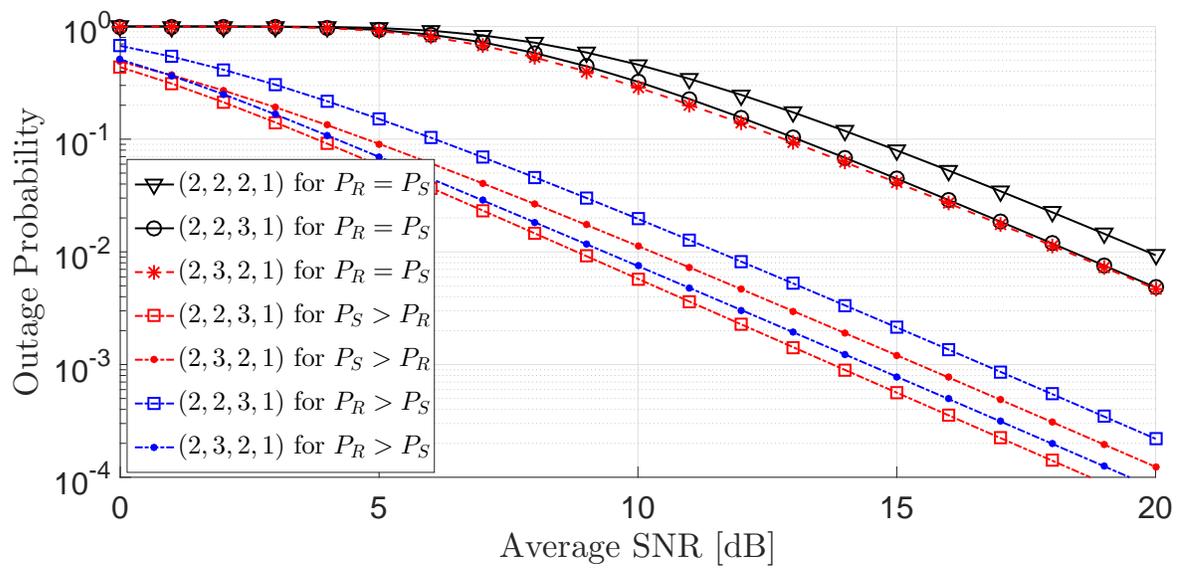}
\caption{Outage probability as the function of average SNR based on single antenna at D with different antenna configurations ($\gamma_T=10$dB and $5$dB for symmetric and asymmetric cases, respectively).} \label{Figure_1}
\end{figure}

Fig.~\ref{Figure_1} shows the results based on the receive ZF with different antenna configurations. We observe some interesting results which are useful for system designers.
Specifically, although there is only single antenna mounted on D, the performance can be improved by applying appropriate design parameters at S and R, especially for receive ZF case. Here, for symmetric case, $(2,3,2,1)$ slightly outperforms $(2,2,3,1)$. 
Note that for asymmetric case, especially for $\text{link}_{RD} > \text{link}_{SR}$ (or  $P'_R:P'_S=3:2$ ) case, this performance gap increases due to the increased possibility of successful decoding at D and the relatively higher diversity order of S-R link for $(2,3,2,1)$ case compared to $(2,2,3,1)$ case.
Contrary, for  $\text{link}_{SR} > \text{link}_{RD}$ (or  $P'_S:P'_R=3:2$ ) case, the possibility of successful decoding at R increases. As a result, $(2,2,3,1)$ with the higher diversity order of R-D link can provide the better performance. 

\begin{figure} 
\centering
\includegraphics[width=7in]{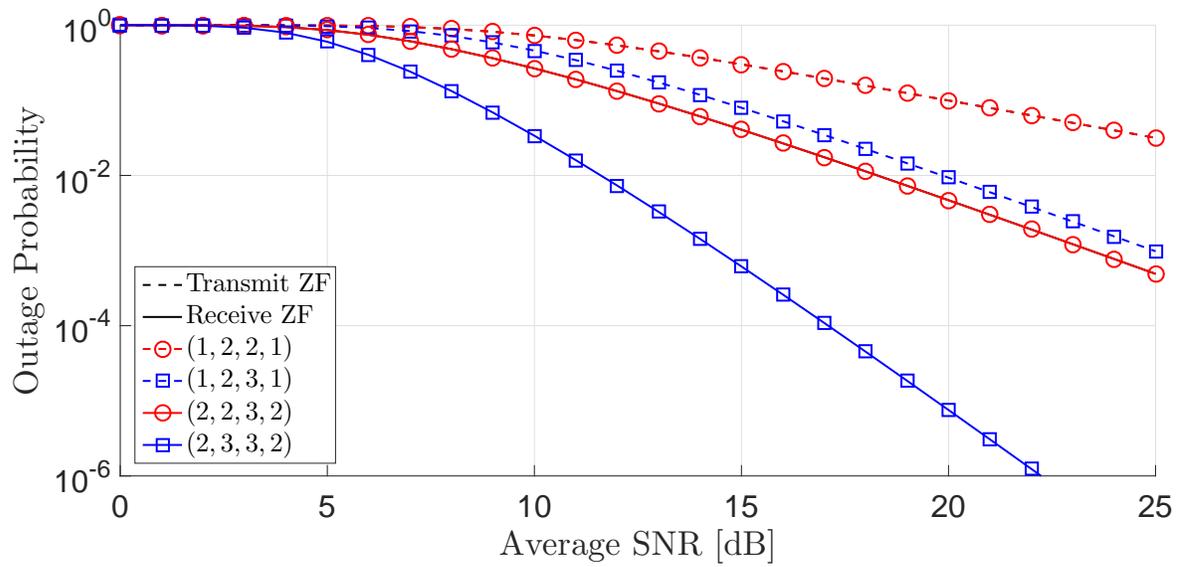}
\caption{Outage probability as the function of average SNR with $\gamma_T=10$dB for both receive and transmit ZFBF cases.} \label{Figure_2}
\end{figure}

In Fig.~\ref{Figure_2}, we observe that an additional performance gain can be obtained via increasing the number of antenna at R. More specifically, for receive ZF case, increasing $N_{R_1}$ at R can obtain the additional performance gain while  for transmit ZF case, the additional performance gain can be obtained via increasing  $N_{R_2}$ .

\begin{figure} 
\centering
\includegraphics[width=7in]{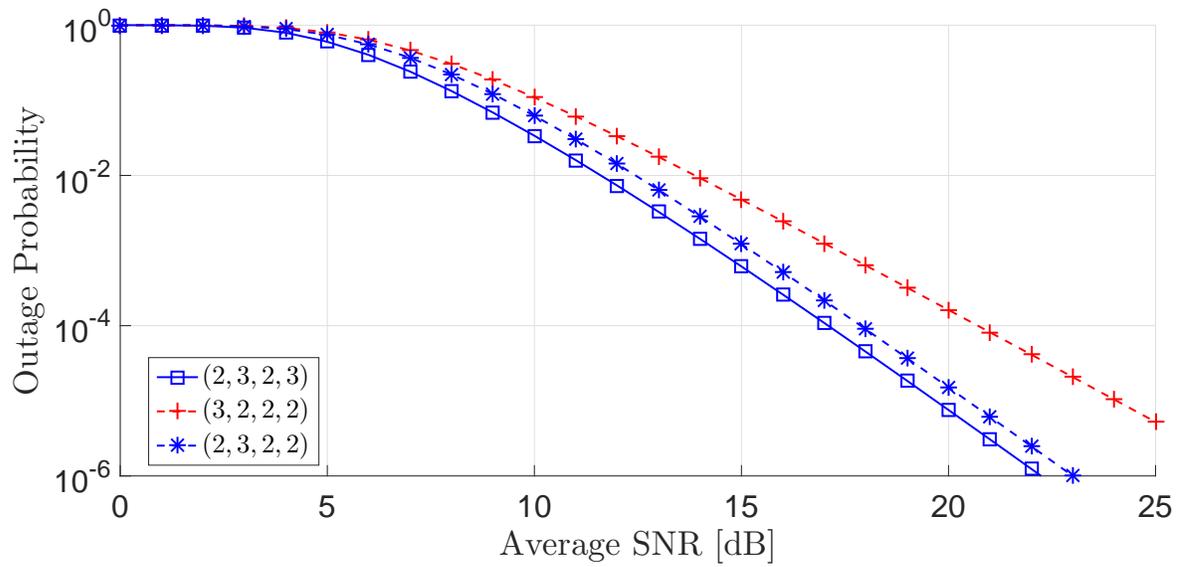}
\caption{Outage probability as the function of average SNR with $\gamma_T=10$dB for receive ZFBF case.} \label{Figure_3}
\end{figure}

In Fig.~\ref{Figure_3}, for both $(2,3,2,2)$ and  $(2,3,2,3)$ cases, the diversity order of these cases is the same. However, the latter case provides the better performance because the latter case has the relatively higher possibility of successful decoding at D. 
Additionally, for $(3, 2, 2, 2)$ and $(2, 3, 2, 2)$ cases, they have the same number of total antenna. However, we observe that, for receive ZF case, swapping  $N_S$  with  $N_{R_1}$  can improve the outage performance.

\appendices
\section{A MATLAB Code for Evaluation of Coefficients, $D_{n,m}$ or $C_{n,m}$ .}\label{AP:1}
\renewcommand{\baselinestretch}{1}
\footnotesize\begin{verbatim}
%%%%%%%%%%%%%%%%%%%%%%%%%%%%%%%%%%%%%%%%%%%%%%%%
% This program is for coefficient calculation. %
%%%%%%%%%%%%%%%%%%%%%%%%%%%%%%%%%%%%%%%%%%%%%%%%
% initialize the values
  N=input('Input the value of N1 = ');
  L=input('Input the value of N2 = ');
  m = min(N1,N2);
  n = max(N1,N2);
% For K_mn Calculation
  a = prod(factorial(m - [1:m]));
  b = prod(factorial(n - [1:m]));
  K_mn = 1/(a*b);
% For G matrix (Symbolic) 
  syms lamda;
  [i_val, j_val] = meshgrid([1:m],[1:m]);
  G = symfun(factorial(n-m+i_val+j_val-2) - igamma(n-m+i_val+j_val-1,lamda),lamda);
% For p_lamda_max (Symbolic)
  p_lamda_max = symfun(K_mn * diff(det(G(lamda)),lamda),lamda);
% Step 1
  f_eq = p_lamda_max;
% Step 2
  k = 1;
  l = n+m-2;
  cont_flag = true;
  A=[];
  while cont_flag
    % Step 3 
      a_kl = double(1/K_mn * limit(f_eq/(lamda^l*exp(-k*lamda)),lamda,+inf));
      a_kl_prime = a_kl * K_mn * factorial(l)/(k^(l+1));
      format rational
      A_tmp=[k l a_kl_prime];
      A=[A;A_tmp];
    % Step 4
      g_eq = K_mn*a_kl*lamda^l*exp(-k*lamda);
      f_eq = f_eq - g_eq;
    % Step 5
      l = l-1;
      if(l >= n-m)
          cont_flag = true;
      else
    % Step 6
          k = k+1;
          if(k <= m)
              l = (n+m-2*k)*k;
          else
              cont_flag = false;
          end;
      end;
  end;
fileID = fopen('Coefficient_Data.dat','w');
fprintf(fileID,'Estimated coefficient values of d_im for N1=%d and N2=%d \r\n', N,L);
fprintf(fileID,'%s      %s        %s \r\n','i','m','d_im');
fprintf(fileID,'---------------------- \r\n');
fprintf(fileID,'%d      %d      %f  \r\n',A');
fclose(fileID);
\end{verbatim}
 \normalsize

\renewcommand{\baselinestretch}{2}

\bibliographystyle{ieeetran}
\bibliography{IEEEabrv,thesis}

\begin{thebibliography}{10}
\providecommand{\url}[1]{#1}
\csname url@rmstyle\endcsname
\providecommand{\newblock}{\relax}
\providecommand{\bibinfo}[2]{#2}
\providecommand\BIBentrySTDinterwordspacing{\spaceskip=0pt\relax}
\providecommand\BIBentryALTinterwordstretchfactor{4}
\providecommand\BIBentryALTinterwordspacing{\spaceskip=\fontdimen2\font plus
\BIBentryALTinterwordstretchfactor\fontdimen3\font minus
  \fontdimen4\font\relax}
\providecommand\BIBforeignlanguage[2]{{%
\expandafter\ifx\csname l@#1\endcsname\relax
\typeout{** WARNING: IEEEtran.bst: No hyphenation pattern has been}%
\typeout{** loaded for the language `#1'. Using the pattern for}%
\typeout{** the default language instead.}%
\else
\language=\csname l@#1\endcsname
\fi
#2}}

\bibitem{Z.Zhang_2015}
Z.~Zhang, X.~Chai, K.~Long, A.~V. Vasilakos, and L.~Hanzo, ``Full duplex
  techniques for 5{G} networks: {S}elf-interference cancellation, protocol
  design, and relay selection,'' \emph{IEEE Commun. Mag.}, vol.~53, no.~5, pp.
  128--137, May 2015.

\bibitem{S.Goyal_2015}
S.~Goyal, P.~Liu, S.~Panwar, R.~Difazio, R.~Yang, and E.~Bala, ``Full duplex
  cellular systems: {W}ill doubling interference prevent doubling capacity?''
  \emph{{IEEE} Commun. Mag.}, vol.~53, no.~5, pp. 121--127, May 2015.

\bibitem{T.Riihonen_2011_2}
T.~Riihonen, S.~Werner, and R.~Wichman, ``Mitigation of loopback
  self-interference in full-duplex {MIMO} relays,'' \emph{{IEEE} Trans. Signal
  Processing}, vol.~59, pp. 5983--5993, Dec. 2011.

\bibitem{B.P.Day_2012}
B.~P. Day, A.~R. Margetts, D.~W. Bliss, and P.~Schniter, ``Full-duplex {MIMO}
  relaying: achievable rates under limited dynamic range,'' \emph{{IEEE} J.
  Select. Areas Commun.}, vol.~30, pp. 1541--1553, Sept. 2012.

\bibitem{Bharadia_2013}
D.~Bharadia, E.~McMilin, and S.~Katti, ``Full duplex radios,'' in \emph{Proc.
  SIGCOMM'13}, Aug. 2013, pp. 375--386.

\bibitem{T.Riihonen_2016}
T.~Riihonen and R.~Wichman, \emph{Full-Duplex in Wireless Communications in
  Wiley Encyclopedia of Electrical and Electronics Engineering}.\hskip 1em plus
  0.5em minus 0.4em\relax John Wiley \& Sons, 2016.

\bibitem{Suraweera_2014}
H.~A. Suraweera, I.~Krikidis, G.~Zheng, C.~Yuen, and P.~J. Smith,
  ``Low-complexity end-to-end performance optimization in {MIMO} full-duplex
  relay systems,'' \emph{{IEEE} Trans. Wireless Commun.}, vol.~13, no.~2, pp.
  913--927, Feb. 2014.

\bibitem{G.Zheng_2015}
G.~Zheng, ``Joint beamforming optimization and power control for full-duplex
  {MIMO} two-way relay channel,'' \emph{IEEE Trans. Signal Process.}, vol.~63,
  no.~3, pp. 555--566, Feb. 2015.

\bibitem{Y.Li_2016}
Y.~Li, N.~Li, M.~Peng, and W.~Wang, ``Relay power control for two-way
  full-duplex amplify-and-forward relay networks,'' \emph{IEEE Signal Process.
  Lett.}, vol.~23, no.~2, pp. 292--296, Feb. 2016.

\bibitem{D.Gunduz_2008}
D.~Gunduz, A.~Goldsmith, and H.~V. Poor, ``Mimo two-way relay channel:
  Diversity-multiplexing tradeoff analysis,'' in \emph{Proc. 42nd Asilomar
  Conf. Signals, Syst. Comput.}, Oct. 2008, pp. 1474--1478.

\bibitem{D.Hwang_2016}
D.~Hwang, Y.~Choi, and T.-J. Lee, ``Optimization of zero forcing beamfomer for
  the full duplex relay system,'' \emph{{IEEE} Trans. Wireless Commun.}, 2016,
  to appear.

\bibitem{K.Wong_2012}
C.~Zhong, T.~Ratnarajah, S.~Jin, and K.~Wong, ``Performance analysis of optimal
  single stream beamforming in {MIMO} dual-hop {AF} systems,'' \emph{{IEEE} J.
  Select. Areas Commun.}, vol.~30, no.~8, pp. 1415--1427, Sept. 2012.

\bibitem{R.Mallik_2003}
P.~Dighe, R.~Mallik, and S.~Jamuar, ``Analysis of transmit-receive diversity in
  {R}ayleigh fading,'' \emph{{IEEE} Trans. Commun.}, vol.~51, no.~4, pp.
  694--703, Apr. 2003.

\bibitem{S.Aissa_2005}
A.~Maaref and S.~Aissa, ``Closed-form expressions for the outage and ergodic
  {S}hannon capacity of {MIMO} {MRC} systems,'' \emph{{IEEE} Trans. Commun.},
  vol.~53, no.~7, pp. 1092--1095, July 2005.

\bibitem{kn:gradshteyn_6}
I.~S. Gradshteyn and I.~M. Ryzhik, \emph{Table of Integrals, Series, and
  Products}, 6th~ed.\hskip 1em plus 0.5em minus 0.4em\relax San Diego, CA:
  Academic Press, 2000.

\bibitem{Q.Zhou_2006}
Q.~Zhou and H.~Dai, ``Asymptotic analysis in {MIMO} {MRT/MRC} systems,''
  \emph{Eur. J. Wireless Commun. Netw.}, vol. 2006, no.~2, pp. 1--8, Apr. 2006.

\end{thebibliography}

\clearpage

\clearpage

\clearpage

\clearpage

%\begin{figure} 
%\centering
%\includegraphics[width=5.5in]{System_figure.eps}
%\caption{System Model}
%\end{figure}

%
%\newpage
%\begin{figure}%[h!]
%\centering
%\includegraphics[width=5.5in]{integration_region.eps}
%\caption{Integral regions for Eq. (\ref{eq:AP-III-1})}
%\label{fig:2}
%\end{figure}

\end{document}